\documentclass{acmsiggraph}

\title{Deep Shading: Convolutional Neural Networks for Screen-Space Shading}

\author{Oliver Nalbach\\Max-Planck-Institut f\"ur Informatik \and Elena Arabadzhiyska\\Max-Planck-Institut f\"ur Informatik \and Dushyant Mehta\\Max-Planck-Institut f\"ur Informatik \and Hans-Peter Seidel\\Max-Planck-Institut f\"ur Informatik \and Tobias Ritschel\\University College London}
\pdfauthor{Oliver Nalbach}


\setcopyright{none}

\copyrightyear{2016}

\usepackage{amsmath}
\usepackage{paralist}
\usepackage{booktabs}
\usepackage{microtype}
\usepackage{amssymb}
\usepackage{soul}
\usepackage{color}
\usepackage{multirow}
\usepackage{tikz}
\usepackage{tabularx}
\usepackage{units}
\usepackage{mathptmx}
\usepackage{mathrsfs}
\usepackage{wrapfig}

\usepackage{pifont}
\newcommand{\cmark}{\ding{51}}%
\newcommand{\xmark}{\ding{55}}%

\newcolumntype{C}[1]{>{\centering}m{#1}}

\newcommand{\eg}{e.\,g.,\ }
\newcommand{\ie}{i.\,e.,\ }
\newcommand{\etal}{~et~al.\ }

\newcommand{\px}[1]{\unit[#1]{px}}

\DeclareGraphicsExtensions{.pdf,.ai,.eps}
\def\figurePath{}

\def\myfigure#1#2{\begin{figure}[ht]\centering\includegraphics*[width = \linewidth]{\figurePath#1}\caption{#2}\label{fig:#1}\end{figure}}

\def\mycfigure#1#2{\begin{figure*}[t]\centering\includegraphics*[clip, width = \linewidth]{\figurePath#1}\caption{#2}\label{fig:#1}\end{figure*}}

\def\mypfigure#1#2{\begin{figure*}[p]\centering\includegraphics*[clip, width = \linewidth]{\figurePath#1}\caption{#2}\label{fig:#1}\end{figure*}}

\def\mysection#1#2{\section{#1}\label{sec:#2}}
\def\mysubsection#1#2{\subsection{#1}\label{sec:#2}}

\newcommand{\refSec}[1]{Sec.~\ref{sec:#1}}
\newcommand{\refFig}[1]{Fig.~\ref{fig:#1}}

\newcommand{\refTbl}[1]{Tbl.~\ref{tbl:#1}}

\begin{document}

\teaser{
  \vspace{-0.5cm}
  \includegraphics[width=\linewidth]{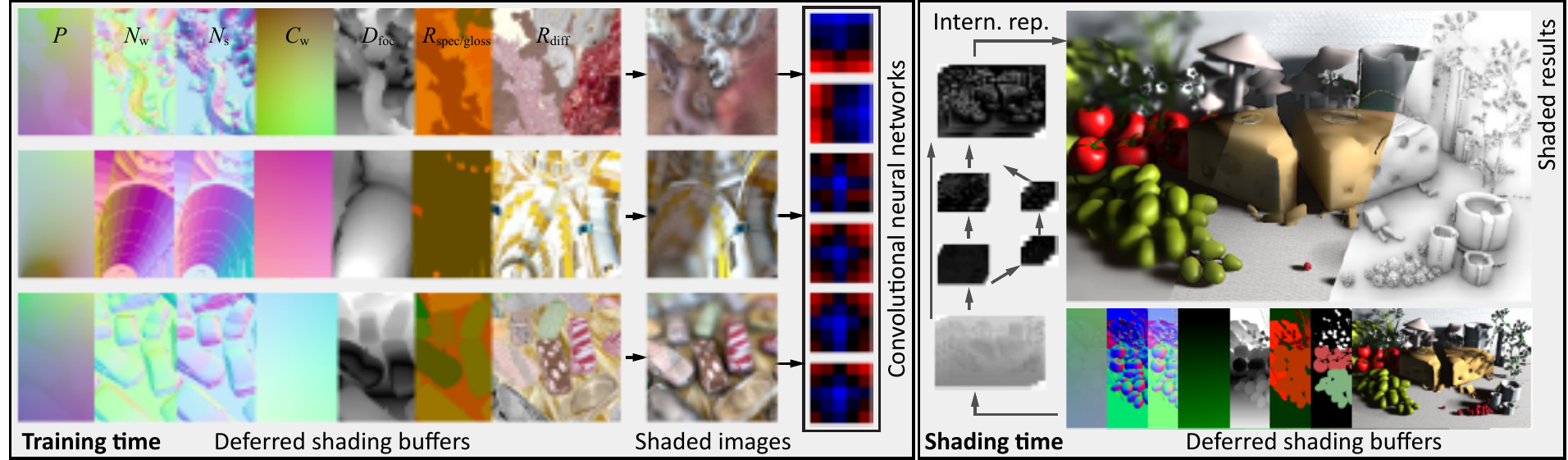}
  \caption{In training \emph{(left)}, our approach learns a mapping from attributes in deferred shading buffers, \eg positions, normals, reflectance, to RGB colors using a convolutional neural network (CNN).
  At run-time \emph{(right)}, the CNN is used to produce effects such as depth-of-field, sub-surface scattering or ambient occlusion at interactive rates (768$\times$512 \px, 1\,ms rasterizing attributes, 21\,/21/\,17\,ms network execution).}
  \label{fig:Teaser}
}

\maketitle

\begin{abstract}
In computer vision, convolutional neural networks (CNNs) have recently achieved new levels of performance for several inverse problems where RGB pixel appearance is mapped to attributes such as positions, normals or reflectance.
In computer graphics, screen-space shading has recently increased the visual quality in interactive image synthesis, where per-pixel attributes such as positions, normals or reflectance of a virtual 3D scene are converted into RGB pixel appearance, enabling effects like ambient occlusion, indirect light, scattering, depth-of-field, motion blur, or anti-aliasing.
In this paper we consider the diagonal problem: synthesizing appearance from given per-pixel attributes using a CNN.
The resulting Deep Shading simulates various screen-space effects at competitive quality and speed while not being programmed by human experts but learned from example images.
\end{abstract}


\keywordlist

\conceptlist

\mysection{Introduction}{Introduction}
The move to deep architectures in machine learning has precipitated unprecedented levels of performance on various computer vision tasks, with several applications having the inverse problem of mapping image pixel RGB appearance to attributes such as positions, normals or reflectance as an intermediate or end objective. Deep architectures have further opened up avenues for several novel applications.
In computer graphics, screen-space shading has been instrumental in increasing the visual quality in interactive image synthesis, employing per-pixel attributes such as positions, normals or reflectance of a virtual 3D scene to render RGB appearance that captures effects such as ambient occlusion (AO), indirect light (GI), sub-surface scattering (SSS), depth-of-field (DOF), motion blur (MB), and anti-aliasing (AA).

In this paper we turn around the typical flow of information through computer vision deep learning pipelines to synthesize appearance from given per-pixel attributes, making use of deep convolutional architectures (CNNs).
We call the resulting approach Deep Shading \cite{nalbach2016deep}.
It can achieve quality and performance similar or better than human-written shaders, by only learning from example data.
This avoids human effort in programming those shaders and ultimately allows to for a deep ``\"ubershader'' that consistently combines all previously separate screen space effects.

\mysection{Previous Work}{PreviousWork}

Previous work comes, on the one hand, from a computer graphics background where attributes have to be converted into appearance and, on the other hand, from a computer vision background where appearance has to be converted into attributes.

\paragraph{Attributes-to-appearance}

The rendering equation \cite{Kajiya1986} is a reliable forward model of appearance in the form of radiance incident at a virtual camera sensor when a three-dimensional description of the scene in form of attributes like positions, normals and reflectance is given.
Several simulation methods for solving it exist, such as finite elements, Monte Carlo path tracing and photon mapping.
The high-quality results these achieve come at the cost of significant computational effort.
Interactive performance is only possible through advanced parallel implementations in specific shader languages \cite{Owens2007}, which not only demands a substantial programming effort, but the proficiency as well.
By choosing to leverage deep learning architectures, we seek to overcome those computational costs by focusing computation on converting attributes into appearance according to example data rather than using physical principles.

Our approach is based on screen-space shading that has been demonstrated to approximate many visual effects at high performance, such as ambient occlusion (AO) \cite{Mittring2007}, indirect light (GI) \cite{Ritschel2009}, sub-surface scattering (SSS) \cite{Jimenez2009}, participating media \cite{Elek2013}, depth-of-field (DOF) \cite{Rokita1993} and motion blur (MB) \cite{McGuire2012}.
Anti-aliasing too can be understood as a special form of screen-space shading, where additional depth information allows to post-blur along the ``correct'' edge to reduce aliasing in FXAA \cite{Lottes2011}.
All of these approaches proceed by transforming a deferred shading buffer \cite{Saito1990}, \ie a dense map of pixel-attributes, into RGB appearance.
We will further show how a single CNN allows combining all of the effects above at once.

Although screen-space shading bears limitations like missing light or shadow from surfaces not part of the image, several properties make it an attractive choice for interactive applications such as computer games: computation is focused only on what is visible on screen; no pre-computations are required making it ideal for rich dynamic worlds; it is independent of the geometric representation, allowing to shade range images or ray-casted iso-surface; it fits the massive fine-grained parallelism of current GPUs and many different effects can be computed from the same input representation.

Until now, image synthesis, in particular in screen-space, has considered the problem from a pure simulation point of view.
In this paper, we demonstrate competitive results achieved by learning from data, mitigating the need for mathematical derivations from first principles.
This has the benefit of avoiding any effort that comes with designing a mathematical simulation model.
All that is required is one general but slow simulation system, such as Monte Carlo, to produce exemplars.
Also, it adapts to the statistics of the visual corpus of our world which might not be congruent to the one a shader programmer assumes.

Applications of machine learning to image synthesis are limited, with a few notable exceptions.
A general overview of how computer graphics could benefit from machine learning, combined with a tutorial from a CG perspective, is given by Hertzmann~\shortcite{Hertzmann2003}.
The CG2Real system \cite{Johnson2011} starts from simulated images that are then augmented by patches of natural images.
It achieves images that are locally very close to real world example data, but it is founded in a simulation system, sharing all its limitations and design effort.
Recently, CNNs were used to transfer artistic style from a corpus of example images to any new exemplar \cite{Gatys2015}.
Our work is different as shading needs to be produced in real-time and in response to a great number of guide signals encoding the scene features instead of just locally changing RGB structures when given other RGB structures.
Dachsbacher \shortcite{Dachsbacher2011} has used neural networks to reason about occluder configurations.
Neural networks have also been used as a basis of pre-computed radiance transfer \cite{Ren2013} (PRT) by running them on existing features to fit a function valid for a single scene.
In a similar spirit, Ren~\etal\shortcite{Ren2015} have applied machine learning to re-lighting: Here an artificial neural network (ANN) learns how image pixels change color in response to modified lighting.
Both works \cite{Ren2013,Ren2015} demonstrate high-qualitfy results when generalizing over light conditions but share the limitation to static 3D scenes, resp.\ 2D images, without showing generalization to new geometry or animations, such as we do.
Such generalization is critical for real applications where geometry is dynamic, resulting in a much more demanding problem that is worth addressing using advanced (\ie deep) learning.
In the same way that PRT was not adopted by the gaming industry for the aforementioned limitations (static geometry), but screen-space shading is, we would argue that only Deep Shading achieves the generalization required to make learning a competitive image-synthesis solution in practice.

Earlier, neural networks were used to learn a mapping from character poses to visibility for PRT \cite{Nowrouzezahrai2009}.
Without the end-to-end learning made possible by deeper architectures, the aforementioned approaches do not achieve generalization between scenes, but remain limited to a specific room, character, etc.
Kalantari\etal\shortcite{Kalantari2015} have used sample data to learn optimal parameters for filtering Monte Carlo Noise.
Our function domain, \ie screen-space attributes, is similar to Kalantari\etal\shortcite{Kalantari2015}.
The range however, is very different.
While they learn filter parameters, we learn the entire shading.
Not much is known about the complexity of the mapping from attributes to filter settings and what is the effect of sub-optimal learning.
In our case, the mapping from attributes to the value is as complex as shading itself.
At the same time, the stakes are high: learning a mapping from attributes to shading results in an entirely different form of interactive image synthesis, not building on anything such as Monte-Carlo ray-tracing that can be slow to compute.
Typically, no end-to-end performance numbers are given for Monte-Carlo noise filtering work such as Kalantari\etal\shortcite{Kalantari2015}.

For image processing, convolution pyramids \cite{Farbman2011} have pursued an approach that optimizes over the space of filters to the end of fast and large convolutions.
Our approach optimizes over pyramidal filters as well, but allows for a much larger number of internal states and much more complex filters defined on much richer input.
Similar to Convolutional Pyramids, our network is based on a ``pyramidal'' CNN, allowing for fast but large filters to produce long-range effects such as distant shadows or strong depth-of-field.

\paragraph{Appearance-to-attributes}
The inverse problem of turning image appearance into semantic and non-semantic attributes lies at the heart of computer vision.
Of late, deep neural networks, particularly CNNs, have shown unprecedented advances in typical inverse problems such as detection \cite{Krizhevsky2012}, segmentation and detection \cite{Girshick2014}, or depth \cite{Eigen2014}, normal \cite{Wang2015} or reflectance estimation \cite{Narihira2015}.
These advances are underpinned by three developments: availability of large training datasets, deep but trainable (convolutional) learning architectures, and GPU accelerated computation.
Another key contributor to these advances has been the ability to train end-to-end, \ie going from input to desired output without having to devise intermediate representations and special processing steps.

One recent advance would be of importance in the application of CNNs to high-quality shading: The ability to produce dense per-pixel output, even for high resolutions, by CNNs that do not only decrease, but also increase resolutions as proposed by \cite{Long2015,Hariharan2015}, resulting in fine per-pixel solutions.
For the problem of segmentation, Ronneberger\etal\shortcite{Ronneberger2015} even apply a fully symmetric U-shaped net where each down-sampling step is matched by a corresponding up-sampling step that may also re-use earlier intermediate results of the same resolution level.

CNNs have also been employed to replace certain graphics pipeline operations such as changing the viewpoint \cite{Dosovitskiy2015,Kulkarni2015}.
Here, appearance is already known, but is manipulated to achieve a novel view.
In our work, we do not seek to change a rendered image but to create full high-quality shading from the basic output of a GPU pipeline such as geometry transformation, visible surface determination, culling, direct light, and shadows.

We seek to circumvent the need to manually concoct and combine convolutions into screen-space shaders that have to be programmed, and ultimately benefit from the tremendous advances in optimizing over deep convolutional networks to achieve a single screen-space \"uber-shader that is optimal in the sense of certain training data.

\mysection{Background}{Background}
Here we briefly summarize some aspects of machine learning, neural networks, deep learning, and training of convolutional networks, to the extent necessary for immediate application to the computer graphics problem of shading.

For our purposes, it suffices to view (supervised) learning as simply fitting a sufficiently complex and high-dimensional function $\tilde f$ to data samples generated from an underlying, unknown function $f$, without letting the peculiarities of the sampling process from being expressed in the fit.
In our case, the domain of $f$ consists of all instances of a per-pixel deferred shading buffer for images of a given resolution (containing per-pixel attributes such as position, normals and material parameters) and the output is the per-pixel RGB image appearance of the same spatial resolution.
We are given the value $f(\mathbf x_i)$ of the function applied to $\mathbf x_i$, the $i$th of $n$ example inputs.
From this we would like to find a good approximation $\tilde f$ to $f$, with the quality of the fit quantified by a cost/loss function that defines some measure of difference between $\tilde f(\mathbf x_i)$ and $f(\mathbf x_i)$.
Training examples can be produced in arbitrary quantity, by mere path tracing or any other sufficiently powerful image synthesis algorithm.

Neural networks (NNs) are a particularly useful way of defining arbitrary non-linear approximations $\tilde f$.
A neural network is typically comprised of computational \emph{units} or \emph{neurons}, each with a set of inputs and a singular scalar output that is a non-linear function of some affine combination of its inputs governed by a vector of \emph{weights} $\mathbf w_k$ for each unit $k$. This affine combination per \emph{unit} is what is learned during training. The \emph{units} are arranged in a hierarchical fashion in \emph{layers}, with the outputs from one \emph{layer} serving as the inputs to the \emph{layers} later in the hierarchy. There usually are no connections between \emph{units} of the same layer.
The \emph{fan-in} of each unit can either connect to all outputs of the previous layer (\emph{fully-connected}), or only \emph{sparsely} to a few, typically nearby ones.
Furthermore, units can also be connected to several preceding layers in the hierarchy.

The non-linearities applied to the affine combination per unit are called \emph{activation functions}.
These are often smooth functions, such as the sigmoid. We make use of \emph{Rectified Linear Units} (ReLUs) which are defined by $r(x) = \max(0, x)$.

Defining $\mathbf{w}$ as the set of weights for the entire network, the function $\tilde f(\mathbf x_i)$ can be expressed as $\tilde f_{\mathbf w}(\mathbf x_i)$. A typical choice of \emph{loss} is the squared $\mathcal L_2$-norm: $|| \tilde f_{\mathbf w}(\mathbf x_i) - f(\mathbf x_i) ||^2_2$.
Alternatively, a perceptual loss function based on a combination of $\mathcal L_1$-norm and structural similarity (SSIM) index may be used \cite{Zhao2015}.
Optimizing weights with respect to the loss is a non-linear optimization process, and \emph{Stochastic Gradient Descent} or its variants are the usual choice of learning algorithm. The method makes a computational time - run time trade-off between computing loss gradients with respect to weights at all exemplars at each gradient descent step and computing gradients with one sample at a particular gradient descent step, by choosing to compute it for subsets of exemplars in mini-batches.
The gradient with respect to $\mathbf w$ is computed by means of \emph{back-propagation}, \ie the error is first computed at the output layer and then propagated backwards through the network \cite{Rumelhart1988}.
From this, the corresponding update to each unit's weight can be computed.

\myfigure{Terminology}{Terminology.}
Convolutional networks are particular neural networks bearing regular spatial arrangement of the units. Within each layer, units are arranged in multiple regular and same-sized grid slices. Each unit in layer $i+1$ connects to the outputs of the units from all slices of layer $i$ within a certain local spatial extent, centered at the unit, defined as the (spatial) kernel size of layer $i+1$. All units of a slice share their weights, \ie the operation of each slice can be seen as a 3D convolution with a kernel as large as the spatial fan-in of the units along two dimensions, and as large as the number of slices in the previous layer along the third dimension. We will refer to \emph{spatial kernel size} simply as \emph{kernel size}.

CNNs typically stack multiple such convolutional layers, with spatial resolution being reduced between consecutive layers as a trick to achieve translation invariance, and computational efficiency for richer features.
However, de-convolutional (or up-sampling) networks allow us to increase the resolution back again \cite{Long2015}, which is critical for our task, where per-pixel appearance \ie high-quality shading needs to be produced quickly.

\mysection{Deep Shading}{DeepShading}

Here, we detail the training data we produced for our task, the network architecture proposed and the process of training it.

\mysubsection{Data Generation}{Training data}

\paragraph{Structure of the Data}

\mycfigure{Scenes}{Selection of images showing training and testing scenes with random textures and lighting.}

Our data sets consist of 61,000 pairs of deferred shading buffers and corresponding shaded reference images in a resolution of $512 \times 512\,px$ for AO and $256 \times 256\,px$ for all other effects.
Of these $61,000$ pairs, we use $54,000$ images to train the network, $6,000$ for validation and the remaining $1,000$ for testing (\refSec{NetworkStructure}).
Train and validation images share the same set of 10 scenes, while the test images come from 4 different scenes not used in training or validation.

To generate the $60,000$ train/validation images, we first render $1,000$ pairs for each of the set of ten scenes of different nature (\refFig{Scenes},\ left part).
These \emph{base images} are then rotated (in steps of $90^\circ$) as well as flipped horizontally and vertically to increase the robustness and size of the training set in an easy way.
Special care has to be taken when transforming attributes stored in view space, here the respective positions and vectors have to be transformed themselves by applying rotations or mirroring.
For the test set, we proceed analogously but using the distinct set of four scenes and appropriately less base images per scene.
Generating one set, \ie rendering and subsequent data augmentation, takes up to about 170 hours of computation on a single high-end GPU.
We plan to make our network definitions and data sets available for use by other research groups.

The base images all show unique and randomly sampled views of the respective scene seen through a perspective camera with a fixed field-of-view of $50^\circ$.
View positions are sampled from a box fitted to the scenes' spatial extents.
\refSec{Results} contains additional information on the training sets for each application.
\refFig{Scenes} shows samples of ground truth image pairs for one instance of the network.

About half of our scenes are common scenes from the computer graphics community such as Crytek Sponza or Sibenik Cathedral and other carefully modeled scenes from sources such as BlendSwap. The remaining scenes were composed by ourselves using objects from publicly available sources to cover as many object categories as possible, \eg vehicles, vegetation or food.
Procedurally generated scenes would be another more sophisticated option.

\paragraph{Attributes}
The deferred shading buffers are computed using OpenGL's rasterization without anti-aliasing of any form.
They contain per-pixel geometry, material and lighting information.
All labels are stored as 16 bit HDR images.

Positions are stored in camera space ($P_\mathrm s$) while normals are stored in camera and world space ($N_\mathrm s$ and $N_\mathrm w$).
Camera space is chosen as, for our shading purposes, absolute world positions do not contain more information than the former and would encourage the network to memorize geometry.
Normals are represented as unit vectors in Cartesian coordinates.
Additionally, depth alone ($D_\mathrm s = P_{\mathrm s, 3}$) and distance to the focal plane ($D_\mathrm{focal}$) are provided to also capture sensor parameters.
To be able to compute view-dependent effects, the normalized direction to the camera ($C_\mathrm w$) and the angle between this direction and the surface normal ($C_\mathrm \alpha$) are additional inputs.

Material parameters ($R$) combine surface and scattering properties.
For surfaces, we use the set of parameters to the Phong~\shortcite{Phong1975} reflection model, \ie RGB diffuse and specular colors (denoted as $R_\mathrm{diff}$ and $R_\mathrm{spec}$) as well as scalar glossiness ($R_\mathrm{gloss}$).
For scattering we use the model by Christensen and Burley \shortcite{Christensen2015} which is parameterized by the length of the mean free path for each color channel ($R_\mathrm{scatt}$).

Direct light (denoted by $L$ or $L_\mathrm {diff}$ for diffuse-only) is not computed by the network but provided as an input to it, as is the case with all corresponding screen-space shaders we are aware of.
Fortunately, it can be quickly computed at run-time and fed into the network.
Specifically, we use the Phong reflection model and shadow maps.

Finally, to support motion blur, per-pixel object motion $F$ is encoded as a 2D polar coordinate in each pixel, assuming that the motion during exposure time is small enough to be approximated well by a translation.
The first component holds the direction between 0 and $\pi$ (motion blur is symmetric), the second component holds the distance in that direction.

In summary, each pixel contains a high-dimensional features vector, where the dimensions are partially redundant and correlated, \eg normals are derivatives of positions and camera space differs from world space only by a linear transformation.
Nonetheless, those attributes are the output of a typical deferred shading pass in a common interactive graphics application, produced within milliseconds from complex geometric models.
Redundant attributes come at almost no additional cost but improve the performance of networks for certain effects.
At the same time, for some effects that do not need certain labels, they can be manually removed to increase speed.

\paragraph{Appearance}
The reference images store per-pixel RGB appearance resulting from shading.
They are produced from virtual scenes using rendering.
Paintings or even real photos would represent valid sample data as well, but their acquisition is significantly more time-consuming than that of the approximate references we use, of which massive amounts can be produced in a reasonable time.

More specifically, we use path tracing for AO, DO and IBL and sample multiple lens positions or points in time for depth-of-field and motion blur, respectively. For anti-aliasing, reference images are computed with $8\times$ super-sampling relative to the label images.

We use 256 samples per pixel (spp) for all of the effects.
While Monte Carlo noise might remain, compute time is better invested into producing a new image.
All per-object attributes which are allowed to vary at run-time (\eg material parameters) are sampled randomly for each training sample.
For effects including depth-of-field and sub-surface scattering we found it beneficial to texture objects by randomly assigned textures from a large representative texture pool \cite{Cimpoi2014} to increase the information content with respect to the underlying blurring operations. 
Automatic per-object box mapping is used to assign UV coordinates.

We do not apply any gamma or tone mapping to our reference images used in training.
It therefore has to be applied as a post-process after executing the network.

In practice, some effects like AO and DO do not compute final appearance in terms of RGB radiance, but rather a quantity which is later multiplied with albedo.
We found networks that do not emulate this obvious multiplication to be substantially more efficient while also requiring less input data and therefore opt for a manual multiplication.
However, the networks for effects that go beyond this simple case need to include the albedo in their input and calculations.
The result section will get back to where albedo is used in detail.
\refTbl{Main} provides an overview in the column ``albedo''.

In a similar vein, we have found that some effects are best trained for a single color channel, while others need to be trained for all channels at the same time.
In the first case, the same network is executed for all three input channels simultaneously using vector arithmetic after training it on scalar images showing only one of the color channels.
In the second case, one network with different weights for the three channels is run.
We refer to the first case as ``mono'' networks, to the latter as ``RGB'' networks (\refTbl{Main}).

\mysubsection{Network}{Network}
\label{Network}
\mycfigure{Network}{\emph{Left:} The big picture with one branch going down and another going up again in a U-shape. 
\emph{Right:}. One level of our network. Boxes represent in- and outputs of the layers, the arrows correspond to the operations performed by the respective layers. The spatial resolution is denoted by multiples of $n$, the number of channels by multiples of $u$.
The convolution groups are not emphasized for simplicity.}

Our network is U-shaped, with a left and a right \emph{branch}.
The first and left branch is reducing spatial resolution (\emph{down} branch) and the second and right branch is increasing it again (\emph{up} branch).
We refer to the layers producing outputs of one resolution as a \emph{level}.
\refFig{Network} shows an example of one such level.
Overall, up to 6 levels with corresponding resolutions ranging from $512\times 512 \px$ to $16\times 16 \px$ are used.
Further, we refer to the layers of a particular level and branch (\ie left or right) as a \emph{step}.
Each step is comprised of a convolution and a subsequent activation layer.
The convolutions (blue in \refFig{Network}) have a fixed extent in the spatial domain, which is the same for all convolutions but may vary for different effects to compute.
Furthermore, we use convolution groups with $2^n$ groups on level $n$. This means that both input and output channels of a convolution layer are grouped into $2^n$ same-sized blocks where outputs from the $m$-th block of output channels may only use values from the $m$-th block of input channels.
The consecutive activation layers (orange in \refFig{Network}) consist of \emph{leaky ReLUs} as described by Maas\etal \shortcite{Maas2013}, which multiply negative values by a small constant instead of zero.

The change in resolution between two steps on different levels is performed by re-sampling layers.
These are realized by $2\times 2$ mean-pooling on the down (red in \refFig{Network}) and by bilinear up-sampling (green in \refFig{Network}) on the up branch.

The layout of this network is the same for all our effects, but the number of kernels on each level and the number of levels vary.
All designs have in common that the number of kernels increases by a factor of two on the down part to decrease by the same factor again on the up part.
We denote the number of kernels used on the first level (\ie level $0$) by $u_0$.
A typical start value is $u_0 = 16$, resulting in a 256-dimensional feature vector for every pixel in the coarsest resolution for the frequent case of 5 levels.
The coarsest level consists of only one step, \ie one convolution and one activation layer, as depicted in \refFig{Network}.
Additionally, the convolution steps in the up-branch access the outputs of the corresponding step of the same output resolution in the down part (gray arrow in \refFig{Network}). This allows to retain fine spatial details.

A typical network has about 130,000 learnable parameters \ie weights and bias terms (see \refTbl{Main} for details).
We call the CNN resulting from training on a specific input and specific labels a \emph{Deep Shader}.

\paragraph{Training}

Caffe \cite{Jia2014}, an open-source neural network implementation, is used to implement and train our networks.
To produce the input to the first step, all input attributes are loaded from image files and their channels are concatenated forming input vectors with $3$ to $18$ components per pixel.
To facilitate learning of networks of varying complexity, without the need of hyper-parameter optimization, particularly of learning rates, we use an adaptive learning rate method (ADADELTA \cite{Zeller2012}) with a \emph{momentum} of $0.9$ which selects the learning rate autonomously.

We use a loss function based on the structural similarity (\emph{SSIM}) index \cite{Zhao2015} which compares two image patches in a perceptually motivated way, and which we found to work best for our task (\refSec{LossComparison}).
The loss between the output of the network and the ground truth is determined by tiling the two images into $8 \times 8 \px$ patches and combining the SSIM values computed between corresponding patches for each channel.
SSIM ranges from $-1$ to $1$, higher values indicating higher similarity. Structural dissimilarity (\emph{DSSIM}) is defined as $(1-SSIM)/2$, and used as the final loss.

\paragraph{Testing}

The test error is computed as the average loss over our test sets (\refSec{Training data}).
The resulting SSIM values are listed in \refTbl{Main}.

\paragraph{Implementation}

While Caffe is useful for training the network, it is inconvenient for use inside an interactive application, \eg the buffers produced by OpenGL would have to be transformed to match Caffe's internal memory layout before being able to execute the network.
Instead of integrating Caffe into a rendering framework we opted to re-implement the forward pass of the network using plain OpenGL shaders operating on array textures.
OpenGL also enables us to use the GPU's hardware support for up- and down-sampling as well as to drop actual concatenation layers by simply accessing two layered inputs instead of one when performing convolutions.
In our application, the Deep Shader output can be interactively explored as seen in the supplemental video.

\mysection{Results}{Results}

\mypfigure{Results}{
Results of different Deep Shaders as discussed in \refSec{Results}.
``Original'' is the input RGB image without any effect.
The lower right panel shows animated variants of the Deep Shader seen above from the supplemental video, running at interactive rates.}

This section analyzes learned Deep Shaders for different shading effects.
\refTbl{Main} provides an overview of their input attributes, structural properties and resulting SSIM achieved on test sets, together with the time needed to execute the network using our implementation on an NVIDIA GeForce GTX 980 Ti GPU.
For visual comparison, we show examples of Deep Shaders applied to new (non-training) scenes compared to the reference implementations used to produce the training sets in \refFig{Results}.

\begin{table}[ht]
	\centering
	\small
	\setlength{\tabcolsep}{3pt}
	\caption{
	Structural properties of the networks for different effects, resulting degrees of freedom, SSIM on the test set and time for executing the network using our OpenGL implementation on $768 \times 512 \px$ inputs. In case of mono networks, the time refers to the simultaneous execution of three networks. The SSIM is always with respect to the raw output of the network, \eg indirect irradiance for GI. The final image might show even better SSIM.
	}
	\begin{tabular}{llccrrrrrr}
Effect & Attributes & Albedo & Mono & $u_0$. & Lev. & Ker. & Size & SSIM & Time \\
	\toprule
IBL
& $N_\mathrm w$, $C_\mathrm \alpha$, $R$		& \cmark & \xmark	& 300		& 1		&$1\times 1$	 & $3.9$ K	& .796 & 28 ms \\
AO
& $N_\mathrm s$, $P_\mathrm s$		& \xmark &	\cmark	&8		&6		&$3\times 3$	 & $71$ K	& .729 & 17 ms \\
DO
& $N_\mathrm w$, $N_\mathrm s$, $P_\mathrm s$ & \xmark &	\xmark	&16		&5		&$3\times 3$	 & $135$ K	&  .589  & 50 ms \\
GI
& $N_\mathrm s$, $P_\mathrm s$, $L_\mathrm {diff}$	& \cmark &	\cmark	&16		&5		& $3\times 3$	 & $134$ K & .798 & 60 ms \\
SSS
& $P_\mathrm s$,  $R_\mathrm{scatt}$, $L$& \cmark &		\cmark	& 8		&5		&$3\times 3$	 & $133$ K & .905 & 21 ms \\
DoF
& $D_\mathrm{focal}$, $L$	& \cmark &	\cmark					&8		&5		& $3\times3$	 & 34 K	&  .848 & 21 ms \\
MB
& $F,L,D_\mathrm s$	& \cmark &	\cmark	& 16		&5		& $3\times 3$	 & $133$ K	& .916 & 58 ms \\
AA
& $D_\mathrm s$, $L$	& \cmark &	\cmark		& 8		&1		&$5\times 5$	 & $1217$ & .982 & 3.8 ms \\
	\midrule
Full
& All & \cmark & \xmark	& 24	&5		& $3\times3$	 & $203$ K	& .667 & 97 ms \\
	\bottomrule
	\end{tabular}
	\label{tbl:Main}
\end{table}

\paragraph*{Ambient Occlusion}

Ambient occlusion, a prototypical screen-space effect, simulates darkening in corners and creases due to a high number of blocked light paths and is typically defined as the percentage of directions in the hemisphere around the surface normal at a point which are not blocked within a certain distance.
Our ground truth images are computed using ray-tracing with a constant effect range defined in world space units.
In an actual application, the AO term is typically multiplied with the ambient lighting term before adding it to the image.

\myfigure{OursVsHBAO}{A same-time comparison between Deep Shading and HBAO. Deep Shading for AO is on-par with state-of-the-art methods like HBAO, both numerically and visually.}

The CNN faithfully reproduces darkening in areas with nearby geometry (\refFig{Results}),
the most noticeable difference to the reference being blurrier fine details.
To evaluate how well our learned shader performs in comparison to optimized screen-space AO techniques, in \refFig{OursVsHBAO}, we show a same-time comparison to Horizon-based Ambient Occlusion (HBAO) \cite{Bavoil2008} which is an efficient technique used in games.
On the test set, HBAO achieves only marginally higher SSIM than our network which we consider remarkable given that our method has been learned by a machine.
We made AO the subject of further in-depth analysis of alternative network designs described in \refSec{Analysis} and seen in \refFig{Loss}, a) and b).

\paragraph*{Image-based Lighting}

In image-based lighting a scene is shaded by sampling directions in a specific environment map to determine incoming radiance, assuming the latter is unblocked.
The network is trained to render a final image based on diffuse and specular colors as well as gloss strengths, so that no further processing is necessary. It operates on all color channels simultaneously.

As can be seen from the vehicles in \refFig{Results}, the network handles different material colors and levels of glossiness well.
The two main limitations are a slight color shift compared to the reference, as seen in the tires of the tractors, and an upper bound on the level of glossiness.
The latter is not surprising as the extreme here is a perfect mirror which would need a complete encoding of the illumination used in training, which has a resolution of several megapixels, into a few hundred network kernels. Generalizing over different environment maps by using the latter as additional network input remains future work.

\paragraph*{Directional Occlusion}

Directional occlusion \cite{Ritschel2009} is a generalization of AO where each sample direction is associated with a radiance sample taken from an environment map and light is summed only from unblocked directions. DO is applied in the same way as AO.
As for AO, ray-tracing is used to resolve occluded directions within a fixed world-space radius.
The training data is produced using one specific environment map, hence the DO Deep Shader, as for IBL, produces only shading for this particular illumination. As IBL it operates on all channels simultaneously.

While AO works well, DO is more challenging for Deep Shading.
The increased difficulty comes from indirect shadows now having different colors and appearing only for certain occlusion directions.
As can be seen in \refFig{Results}, the color of the light from the environment map and the color of shadows match the reference but occlusion is weakened in several places.
This is due to the fact that the indirect shadows resulting from DO induce much higher frequencies than unshadowed illumination or the indirect shadows in AO, which assume a constant white illumination from all directions, and are harder to encode in a network.

\paragraph*{Diffuse Indirect Light}

A common challenge in rasterization-based real-time rendering is indirect lighting.
To simplify the problem, the set of relevant light paths is often reduced to a single ``indirect bounce'', diffuse reflection \cite{Tabellion2004} and restricted to a certain radius of influence.
The ground truth in our case consists of the ``indirect radiance'', \ie the light arriving at each pixel after one interaction with a surface in the scene.
From this, the final indirect component can be computed by multiplying with the diffuse color.
We compute our ground truth images in screen-space.
The position of the light source is sampled uniformly at random per image.
As we are assuming diffuse reflections, the direct light input to the network is computed using only the diffuse reflectance of the material.
In the absence of advanced effects like fluorescence or dispersion, the light transport in different color channels is independent from each other. We therefore apply a monochromatic network.
The network successfully learns to brighten areas in shadow, which do not appear pitch-black anymore, rather the color of nearby lit objects (\refFig{Results}).

\paragraph*{Anti-aliasing}

While aliasing on textures can be reduced by applying proper pre-filtering, this is not possible for sharp features produced by the geometry of a scene itself.
Classic approaches compute several samples of radiance per pixel which typically comes with a linear increase in computation time.
This is why state-of-the-art applications like computer games offer simple post-processing filters like fast approximate anti-aliasing (FXAA) \cite{Lottes2011} as an alternative, which operate on the original image and auxiliary information such as depth values.
We let our network learn such a filter on its own, independently for each channel.

Applying our network to an aliased image (\refFig{Results}) replaces jagged edges by smooth ones. While it cannot be expected to reach the same performance as the 8$\times$ multi-sample anti-aliasing (MSAA) we use for our reference, which can draw from orders of magnitude of additional information, the post-processed image shows fewer disturbing artifacts.
At the same time, the network learns to not over-blur interior texture areas that are properly sampled, but only blurs along depth discontinuities.

\paragraph*{Depth-of-field}

As a simple rasterization pass can only simulate a pinhole camera, the appearance of a shallow depth of field (DoF) has to be faked by post-processing when multiple rendering passes are too costly.
In interactive applications, this is typically done by adaptive blurring of the sharp pinhole-camera image.
We learn our own depth-of-field blur from sample data which we generate in an unbiased way, by averaging renderings from multiple positions on the virtual camera lens.
The amount of blurriness depends on the distance of each point to the focal plane.
As the computation of the latter does not come with any additional effort compared to the computation of simple depth, we directly use it as an input to the Deep Shader.
While the training data is computed using a fixed aperture, the shallowness of the depth of field, as well as the focusing distance, are easy to adjust later on by simply scaling and translating the distance input.
The Deep Shader again is trained independently for each channel, assuming a non-dispersive lens.

The Deep DoF Shader blurs things in increasing distance from the focal plane by increasing extents.
In \refFig{Results}, the blossoms appear sharper than \eg leaves in the background.
It proved fruitful to use textured objects in training to achieve a sufficient level of sharpness in the in-focus areas.

\paragraph*{Sub-surface Scattering}
Simulating the scattering of light inside an object is crucial for achieving realistic appearance for translucent materials like wax and skin.
A popular approximation to this is screen-space sub-surface scattering (SSSS) \cite{Jimenez2009} which essentially applies a spatially-varying blurring kernel to the different color channels of the image.
We produce training data at every pixel by iterating over all other pixels and applying Pixar's scattering profile \cite{Christensen2015} depending on the distance between the 3D position at the two pixels.
After training the Deep Shader independently for all RGB channels on randomly textured training images with random parameters to the blurring profile we achieve images which are almost indistinguishable from the reference method.

\paragraph*{Motion Blur}
Motion blur is the analog to depth-of-field in the temporal domain.
Images of objects moving with respect to the camera appear to be blurred along the motion trajectories of the objects for non-infinitesimal exposure times.
The direction and strength of the blur depends on the speed of the object in the image plane  \cite{McGuire2012}.

For training, we randomly move objects inside the scene for random distances.
Motions are restricted to those which are parallel to the image plane, so that the motion can be encoded by an angle and magnitude alone.
We also provide the Deep Shader with a depth image to allow it to account for occlusion relations between different objects correctly, if possible.
Our Deep Shader performs motion blur in an convincing way that manages to convey a sense of movement and comes close to the reference image (\refFig{Results}).

\paragraph*{Full Shading}

Finally, we learn a Deep Shader that combines several shading effects at once and computes a scene shaded using image-based lighting, with ambient occlusion to produce soft shadows, and additional shallow depth-of-field.
As AO and IBL are part of the effect, the network can again make use of all channels simultaneously and is again specific to a certain environment map.
Note, that a single Deep Shader realizes all effects together in a single network. 

An image generated using the network (\refFig{Results}) exhibits all of the effects present in the training data.
The scene is shaded according to the environment map, working for both diffuse and moderately glossy materials.
Furthermore, there is a subtle depth-of-field effect.

\paragraph*{Animations}
Please see the supplemental video for view changes inside those scenes, and dynamic characters.

\mysection{Analysis}{Analysis}

In the first part of this section, we address some shortcomings in the form of typical artifacts produced by our method and also discuss how the network reacts when applying it to new resolutions and attributes rendered with a different field-of-view value.
The remainder of the section explores some of the countless alternative ways to apply CNNs, and machine learning in general, to the problem of screen-space shading. We cover different choices of actual network structure (\refSec{NetworkStructure}), loss function (\refSec{LossComparison}) and training data anatomy (\refSec{TrainingData}) as well two techniques competing with deep CNNs, namely artificial neural networks (ANNs) and random forest (RFs) (\refSec{RegressionComparison}).

\mysubsection{Visual Analysis}{VisualAnalysis}

\paragraph*{Typical Artifacts}

\myfigure{Artifacts}{
Typical artifacts of our approach: 
\emph{a):} Blur.
\emph{b):} Color shift.
\emph{c):} Ringing.
\emph{d):} Background darkening.
\emph{e):} Attribute discontinuities.
}

In networks, where light transport becomes too complex and the mapping was not fully captured, what looks plausible in a static image may start to look wrong in a way that is hard to compare to common errors in computer graphics: spatio-temporal patterns resembling the correct patterns emerge, but are inconsistent with the laws of optics and with each other, adding a painterly and surrealistic touch.
We show exemplary artifacts in \refFig{Artifacts}.
Capturing high frequencies is a key challenge for Deep Shaders (\refFig{Artifacts}, a).
If the network does not have enough capacity or was not train enough the results might over-blur with respect to the reference.
We consider this a graceful degradation compared to typical artifacts of man-made shaders such as ringing or Monte Carlo noise which are unstable over time and unnatural with respect to natural image statistics.
Sometimes, networks trained on RGB tend to produce color shifts (\refFig{Artifacts}, b).
CNN-learned filters may also introduce high frequencies manifesting as ringing (\refFig{Artifacts}, c).
Sometimes effects propagate into the wrong direction in world space, \eg geometry may cast occlusions on things behind it (\refFig{Artifacts}, d).
At attribute discontinuities, the SSIM loss lacking an inter-channel prior gives rise to color ringing (\refFig{Artifacts}, e).

\paragraph*{Effect Radius}

\myfigure{SpatialScale}{Increasing (decreasing) the resolution shrinks (enlarges) effect size relative to the resolution. 
The radius can be increased (decreased) again with no effects on timings by scaling the input attribute determining the radius accordingly, \eg the positions for AO.
Consequently, images on the diagonal are similar.
All images show outputs produced by the same network.
The time for each row is identical.
The SSIM is higher for smaller effect radii that are easier to reproduce.}

Typically, screen-space shading is faded out based on a distance term and only accounts for a limited spatial neighborhood.
As we train in one resolution but later apply the same trained network also to different resolutions, the effective size of the neighborhood changes.
As a solution, when applying the network at a resolution which is larger by factor of $N$ compared to the training resolution, we also scale the effect radius accordingly, dividing it by $N$.
While the effect radius is not an input to the network but fixed in the training data, this can still be achieved by scaling the attributes determining the spatial scale of the effect, \eg of the camera space positions in the case of AO, DO or GI, or of the distance to the focal plane in the case of DoF.
To conclude, effect radius and resolution can be changed at virtually no additional cost without re-training the network.

\paragraph*{Internal Camera Parameters}

\myfigure{FOVEffect}{Effect of FOV on image quality.
The horizontal axis is FOV in degrees.
The central line is the reference of 50$^\circ$.
The vertical axis is DSSIM error (less is better).
Note that the vertical axis spans only a small difference ($.12$ to $.18$), indicating FOV has no large impact on visual quality.}

As deep shading was trained on images of a specific FOV of 50$^\circ$, it is not clear how they perform on frame-buffers produced using a different FOV.
\refFig{FOVEffect} demonstrated the effect of FOV on image quality.
We see from the minimal drop in SSIM  quality, that FOV affects the quality only to a very limited extent.

\mysubsection{Network Structure}{NetworkStructure}

\mycfigure{Loss}{
Analysis of different network structures.
We here compare different design choices for different effects in terms of compute time and DSSIM loss.
The vertical axes on all plots corresponds to DSSIM loss (less is better).
The horizontal axes of the line plots range over the number of training iterations.
The scatter plots have computation time of the Deep Shader as the horizontal axis.
\emph{a)} Train, test and validation loss as a function of iterations for different designs of AO (curves).
\emph{b)} Relation of final loss and compute time for different designs for AO.
\emph{c)} Loss as a function of iterations for the chosen designs for other effects (curves).
\emph{d)} Comparison of compute time and final loss for the other effects, as a means of placing their relative complexity.
}

Deep learning architectures, with their vast number of trainable parameters, tend to over-fit even in the presence of a large training corpus. While this typically falls under the purview of regularization, of concern to us is the trade-off between the expressiveness of the network in approximating a certain effect and its computational demand.
To understand this, we investigate two modes of variation of the number of parameters of the network, choosing to vary the spatial extent of the kernels as well as the number of kernels on the first level $u_0$, which also determines the number of kernels for the remaining levels. (See \ref{Network} for details)
We seek the smallest network with adequate learning capacity, that generalizes well on previously unseen data.
The results are summarized in \refFig{Loss},\,(a),\,(b) for the example of AO.

\paragraph*{Spatial Kernel Size}

\refFig{Loss},\,(a) (green and yellow lines) shows the evolution of training, validation and test error with an increasing number of training iterations, with the number of kernels fixed to a medium value of $u_0 = 8$ and varying the spatial extent of the kernels.

We see that with a kernel size of $5 \times 5$, the training profile slightly lags behind that for kernel size of $3 \times 3$, even though they approach a similar test loss at 100k iterations. This shows that both networks have sufficient capacity to approximate the mapping, with neither beginning to overfit. Looking at the run times, however, we see that the one with a kernel size of $3\times3$ is about twice as fast as the one with $5\times5$. This also has a proportional bearing on the training time, given similar mini-batch sizes, \ie the time it takes per iteration.  
Thus, for the given training set, $3 \times 3$ is the optimal choice and is the faster-to-execute of the options as shown in \refFig{Loss},\,(b).
We observe a similar relative timing relationship between the pairs of networks with $u_0 = 4$ and $u_0 = 16$.

\paragraph*{Initial Number of Kernels}
The orthogonal mode of variation is $u_0$, the number of kernels on the first level, with the number of kernels in subsequent layers expressed as multiples of $u_0$.
Again, we plot the training, validation and test errors, this time for different values of $u_0$ (\refFig{Loss},\,a, green and blue lines, yellow and purple lines).
We can observe that reducing the number to $u_0 = 4$ clearly evinces a loss of expressiveness, evidenced by both the training curves as well as the test losses.

Further, nets with $u_0 = 16$ performs only slightly better than $u_0 = 8$ (\refFig{Loss},\,b,), but lose out in compute time by more than a factor of $6$, in part due to increased memory consumption, both in the way of increased number of parameters and increased size of intermediate representations. Varying the spatial kernel size in isolation does not affect the size of intermediate representations but only the number of parameters, which is relatively insignificant compared to the combined memory usage of the intermediate representations.

The number of iterations shown, though sufficient to already make a decision about the choice of structural parameters, still leave the network with scope to learn more (indicated by the negative slope on the train-test curves). Once $u_0 = 8$ with a kernel size of $3\times3$ emerges as the clear choice for our application, we let it train for an additional 100k iterations, keeping an eye on the validation curve for signs of overfitting.

\paragraph*{Structural Choices for other Effects}
The detailed analysis for AO yields an expedient direction to proceed in for the choice of kernel size and $u_0$ for the other effects. We start off with spatial extents of $3 \times 3$ and $5 \times 5$, with $u_0 = 8$, and proceed to increase or decrease $u_0$ in accordance with over-fit\,/\,underfit characteristics exhibited by the the train-test error curves. \refTbl{Main} indicates the final choice of the network structure for each effect. Additionally, the train-test error curves for the final choices for each effect are shown in \refFig{Loss},\,(c), with their test loss-vs.-speed characteristics captured in \refFig{Loss},\,(d). $u_0$ for all pairs of curves in \refFig{Loss},\,(c) are as listed in \refTbl{Main}.

\mysubsection{Choice of Loss Function}{LossComparison}

\myfigure{LossComparison}{Outputs produced by the same network trained with different loss functions for the case of AO.}

The choice of loss function in the optimization has a significant impact on how Deep Shading will be perceived by a human observer.
We trained the same network structure using the common L1 and L2 losses as well as the perceptual SSIM metric and also using combinations of the three. \refFig{LossComparison} shows a visual comparison of results produced by the respective nets.
We found L1 and L2 to be prone to producing halos instead of fading effects out smoothly as can be seen in the first two columns. The combination of L2 with SSIM also exhibits these kind of artifacts to a lesser extent. SSIM and SSIM + L1 both produce visually pleasing results with pure SSIM being more faithful to the amount of contrast found in the reference images.

\mysubsection{Training Data Diversity}{TrainingData}

\myfigure{TrainingDataTradeoffs}{\emph{Left:} Data points correspond to the same time budget to produce training data but with different trade-offs regarding the scene count. \emph{Right:} Patches from a test scene.
}

Ideally, the training set for an effect consists of a vast collection of images from a high number of different scenes with no imperfections from Monte Carlo noise.
However, in practice, the time budget to produce training data is typically limited and the question how to spend this time best arises.
One factor that has influence on the quality of the trained Deep Shaders is the diversity of scenes in the training set, \eg a CNN that has only seen round objects during training will fail to correctly re-produce its effect for square objects.
In our training sets, we use 1000 views from each of 10 different scenes as our starting points (cf.\ \refSec{TrainingData}). To see how well CNNs perform for less diverse data we produced DO training sets of the same total size but for a lower number of different scenes. DO was chosen as we observed it to be particularly sensitive to the scene diversity. The resulting DSSIM values for (the same) test set are plotted in \refFig{TrainingDataTradeoffs},\ left. While the error for five scenes compared to a single one is $5\%$ smaller, increasing the number further to 10 scenes leads to only a smaller advantage of about another $1 \%$ which indicates that our scene set is of acceptable diversity. In the case of DO, the difference in the loss visually translates to a more correct placement of darkening. A network trained with only one scene tends to create ``phantom occlusions'' in free spaces (\refFig{TrainingDataTradeoffs},\ right).

\mysubsection{Comparison With Other Regression Techniques}{RegressionComparison}

Aside from deep CNNs, approaches such as shallow artificial neural networks (ANNs) and random forests (RFs) \cite{Criminisi2013} could putatively be used for our objective, having found use in image synthesis related tasks such as estimation of filter parameters \cite{Kalantari2015} or relighting \cite{Ren2013,Ren2015}.

For comparison, we train our Deep Shader, ANNs, and RFs to regress AO on patches of deferred shading information with a spatial resolution of $256\times256$. For RFs, we input patch sizes of $21\times21$ and predict AO of the patch's central pixel with different number of trees. The RFs are split across four cores and have a minimum of five samples per leaf. With ANNs, we input patch sizes of $11\times11$ and $21\times21$ to predict AO of the central pixel, with 2 hidden layers with 50 nodes each.

The RFs and ANNs are trained on $\mathcal L_2$ loss as in \cite{Ren2013,Ren2015}, and evaluated on the SSIM loss we suggest. For ANNs, speed is measured using the OpenGL implementation of the forward pass of the network (as with CNNs), and for RFs we employ scikit-learn \cite{scikit-learn} on a regular workstation.We see in  \refFig{ANNAndRFComparison} that image quality and run-time depend on the chosen patch size and the number of trees.

\myfigure{ANNAndRFComparison}{
Comparison of AO computed using random forests, shallow neural networks and our Deep Shader.
The vertical axis is image error indicated as DSSIM (less is better) on a linear scale.
The horizontal axis is compute time for $256\times256$ images on a logarithmic scale (less is better). \emph{n} indicates the number of trees.}

\refFig{ANNAndRFComparison}, shows the relative speed and quality for ANNs and RFs compared to Deep Shading. Pixel-wise predictions with random forests clearly lose out on both visual quality and run time. RF run times increase linearly with the number of trees, more of which are necessary to construct a better ensemble. Even with more readily prallelizable variants of RFs \cite{bosch2007image}, there would have to be an improvement of more than two orders of magnitude on the run time to be comparable with our Deep Shader. Besides, there is still the question of image quality. Some structural information captured via multi-variate regression of smaller patches (rather than pixel predictions) may see improvements on this front, and so would increasing the patch size, but again at the cost of run time.

For the two ANNs with patches of size $11\times11$ and $21\times21$, we observe a worsening of image quality with increased patch sizes due to the overfitting owing to the quadratic increase in the number of parameters with patch size, necessitating far more training data and training iterations.

This is a deciding factor in choosing deeper convolutional networks, to allow for an increase in effective receptive field sizes through stacked convolutions and downsampling, without an exponential increase in the number of parameters, while leveraging the expressive power of deeper representations \cite{Hastad1986almost}.

\mysection{Conclusion}{Conclusion}

We propose Deep Shading, a system to perform shading using CNNs.
In a deviation from previous applications in computer vision using appearance to infer attributes, Deep Shading leverages deep learning to turn attributes of virtual 3D scenes into appearance.
It is also the first example of performing complex shading purely by learning from data and removing all considerations of light transport simulation derived from first principles of optics.

We have shown that CNNs can actually model any screen-space shading effect such as ambient occlusion, indirect light, scattering, depth-of-field, motion blur, and anti-aliasing, as well as arbitrary combinations of them at competitive quality and speed.
Our main result is a proof-of-concept of image synthesis that is not programmed by human experts but learned from data without human intervention.

The main limitation of Deep Shading is the one inherent to all screen-space shading techniques, namely missing shading from objects not contained in the image due to occlusion, clipping or culling.
At the same time, screen-space shading is well-established in the industry due to its ability to handle large and dynamic scenes in an output-sensitive manner.
We would also hope that in future refinements, the Deep Shader could even learn to fill in this information, \eg it might recognize the front of a sphere and know that in a natural scene the sphere will have a symmetric back that will cast a certain shadow.
In future work, we would like to overcome the limitation to screen space effects by working on a different scene representation, such as surfels, patches or directly in the domain of light paths.
Some shading effects like directional occlusion and indirect lighting are due to very complex relations between screen space attributes.
Consequently, not all configurations are resolved correctly by a network with limited capacity, such as ours which runs at interactive rates.
We have however observed that the typical artifacts are much more pleasant than from human-designed shaders.
Typical ringing and over-shooting often produces patterns the network has learned from similar configurations, and what appears plausible to the network is often visually plausible as well.
A perceptual study could look into the question whether Deep Shaders, in addition to their capability to learn shading, also produce more visually plausible errors than the typical simulation-type errors which are patterns that never occur in the data.
Screen-space excels in handling complex dynamic scenes, and Deep Shading does as well.
Deep Shaders that result in a low final test error (\refFig{Loss}, c) are almost free of temporal artifacts as seen in the supplemental video.

Deep Shading of multiple effects can currently achieve performance on-par with human-written code, but not exceed it.
We would hope that more and diverse training data, advances in learning methods, and new types of deep representations or losses, such as network losses \cite{Johnson2016} will allow surpassing human shader programmer performance in a not-so-distant future.

\bibliographystyle{acmsiggraph}

\bibliography{article}
\end{document}